\documentclass[aps,prb,twocolumn,noeprint,citeautoscript,superscriptaddress]{revtex4-1}
\usepackage{cancel}
\usepackage{amsmath}
\usepackage{physics}
\usepackage{graphicx}
\usepackage{amssymb}
\usepackage{amsthm}
\usepackage{xspace}
\usepackage{bm}

\usepackage{threeparttable} % Table with notes
\usepackage{dcolumn}
	\newcolumntype{d}[1]{D{.}{.}{#1}}
\usepackage{booktabs} % To thicken table lines

\usepackage{braket}% Enables braket notation

\usepackage{ragged2e}
\usepackage{txfonts}

\usepackage{color}
\usepackage[usenames,dvipsnames]{xcolor}
\usepackage{colortbl}
    \definecolor{myblue}{rgb}{0,0,1}
\usepackage[breaklinks=true,colorlinks=true,linkcolor=myblue,urlcolor=myblue,citecolor=myblue]{hyperref}
\usepackage{cleveref}	
	\crefname{figure}{Figure}{Figures}
	\crefname{table}{Table}{Tables}
	\crefname{equation}{Eq.}{Eqs.}
	\crefname{section}{Section}{Sections}
	\crefname{subsection}{Section}{Sections}
	\crefname{subsubsection}{Section}{Sections}
	\crefname{algorithm}{Algorithm}{Algorithms}
	\creflabelformat{equation}{#2#1#3}
	\crefrangelabelformat{equation}{#3#1#4 to~#5#2#6}

\let\vr\undefined
\newcommand{\vr}{{\bm{r}}}

\newcommand{\vq}{{\bm{q}}}
\newcommand{\vk}{{\bm{k}}}
\newcommand{\vG}{{\bm{G}}}

\newcommand{\vT}{{\bm{T}}}

\newcommand{\p}[1]{{#1}\vk_{#1}}

\newcommand{\cct}[2]{t_{\p{#1}}^{\p{#2}}}
\newcommand{\cctt}[4]{t_{\p{#1}\p{#2}}^{\p{#3}\p{#4}}}
\newcommand{\ccr}[2]{r_{\p{#1}}^{\p{#2}}}
\newcommand{\ccrr}[4]{r_{\p{#1}\p{#2}}^{\p{#3}\p{#4}}}

\begin{document}

\title{Excitons in solids from periodic equation-of-motion coupled-cluster theory}

\author{Xiao Wang}
\affiliation{Center for Computational Quantum Physics, Flatiron Institute, New York, New York 10010 USA}
\author{Timothy C. Berkelbach}
\email{tim.berkelbach@gmail.com}
\affiliation{Center for Computational Quantum Physics, Flatiron Institute, New York, New York 10010 USA}
\affiliation{Department of Chemistry, Columbia University, New York, New York 10027 USA}

\begin{abstract}
We present an ab initio study of electronically excited states of
three-dimensional solids using Gaussian-based periodic equation-of-motion
coupled-cluster theory with single and double excitations (EOM-CCSD).  The
explicit use of translational symmetry, as implemented via Brillouin zone
sampling and momentum conservation, is responsible for a large reduction in
cost.  Our largest system studied, which samples the Brillouin zone using 64
$k$-points (a $4\times 4\times 4$ mesh) corresponds to a canonical EOM-CCSD
calculation of 768 electrons in 640 orbitals.  We study eight simple
semiconductors and insulators, with direct singlet excitation energies in the
range of 3 to 15~eV.  Our predicted excitation energies exhibit a mean absolute
error of 0.27~eV when compared to experiment.  We furthermore calculate the
energy of excitons with nonzero momentum and compare the exciton dispersion of
LiF with experimental data from inelastic X-ray scattering.  By calculating
excitation energies under strain, we extract hydrostatic deformation potentials
in order to quantify the strength of interactions between excitons and acoustic
phonons.  Our results indicate that coupled-cluster theory is a promising method
for the accurate study of a variety of exciton phenomena in solids.
\end{abstract}

\maketitle

\section{Introduction}
\label{sec:intro}

Density functional theory (DFT) is the workhorse of computational materials
science and commonly predicts ground-state properties with high
accuracy~\cite{kohn1965,kresse1996,staroverov2004,paier2006,haas2009,norskov2011,burke2012,freysoldt2014}.
Extending this success to the prediction of excited-state properties is an
ongoing challenge.
Time-dependent DFT (TDDFT)~\cite{runge1984} struggles to treat neutral
excitations in the condensed phase and reasonable accuracy requires the use of
hybrid functionals and/or frequency-dependent exchange-correlation
kernels~\cite{gavrilenko1997,tokatly2001,reining2002,onida2002,sottile2005,botti2007,paier2008,izmaylov2008,sharma2011,rigamonti2015,ullrich2016}.
Instead, the community typically builds upon DFT through Green's function-based
many-body perturbation theory~\cite{onida2002}.  For
weakly-correlated materials, including simple metals, semiconductors, and
insulators, the GW approximation to the
self-energy~\cite{hedin1965,strinati1980,strinati1982,hybertsen1985,hybertsen1986,pulci2005}
combined with the same approximation to the Bethe-Salpeter
equation~\cite{albrecht1998,hanke1980,sham1966,strinati1984,rohlfing2000} yields
excited state properties that are typically in good agreement with experiment.
Strongly-correlated materials are more commonly treated via dynamical mean-field
theory (DMFT)~\cite{georges1992,georges1996}, either in the
DFT+DMFT framework~\cite{georges2004,kotliar2006,held2007} or the GW+DMFT
framework~\cite{biermann2003,biermann2014}.
Alternatively, quantum Monte Carlo methods have been adapted for the
calculation of excitation energies, including variational~\cite{zhao2019} and diffusion Monte 
Carlo~\cite{williamson1998,hunt2018,yang2019qmc} and auxiliary-field quantum Monte Carlo~\cite{ma2013}.

In recent years, wavefunction-based techniques from the quantum chemistry
community have been adapted for periodic boundary conditions and applied to
condensed-phase systems.  For ground-state properties, we mention the
application of M\o ller-Plesset perturbation theory~\cite{sun1996,ayala2001,pisani2005,marsman2009}, coupled-cluster
theory~\cite{hirata2001,hirata2004,katagiri2005,gruneis2011,mcclain2017}, and -- for small supercells -- full configuration interaction
quantum Monte Carlo~\cite{booth2013}.  Charged excitation energies, as quantified
through the one-particle spectral function or the band structure, have been
studied by periodic equation-of-motion coupled-cluster theory for the uniform
electron gas~\cite{mcclain2016}, for the simple semiconductors silicon and diamond~\cite{mcclain2017},
for two-dimensional MoS$_2$~\cite{pulkin2019}, and for transition metal oxides~\cite{gao2019}.

Neutral excitation energies have been primarily studied by configuration
interaction with single excitations (CIS)~\cite{hirata1999,lorenz2011}, which is
equivalent to the use of the Hartree-Fock self-energy in the Bethe-Salpeter
equation (with the Tamm-Dancoff approximation).  Due to the unscreened nature of
the Coulomb interaction, periodic CIS typically predicts excitation energies
which are much too large~\cite{lorenz2012}.  Recently, periodic
equation-of-motion coupled-cluster theory with single and double excitations
(EOM-CCSD) was applied to one-dimensional polyethylene in a small single-particle
basis set~\cite{katagiri2005}, producing results in reasonable agreement with
experiment.  For three-dimensional condensed-phase systems, our group has 
recently applied EOM-CCSD to the neutral excited-state properties of the uniform
electron gas~\cite{lewis2019} as well as the absorption and pump-probe
spectroscopy of the naphthalene crystal at the $\Gamma$ point~\cite{lewis2019a}. 

Here, we continue this endeavor and present the results of a new EOM-CCSD
implementation with translational symmetry and Brillouin zone sampling for
three-dimensional atomistic solids.  The layout of the article is as follows. In
section \ref{sec:theory}, we describe the theory underlying our implementation,
including details about our symmetry-adapted Gaussian basis sets and periodic EOM-CCSD.  
In section~\ref{sec:comput}, we present
computational details, including details about the materials studied, the basis
sets used, and integral evaluation.  In section~\ref{sec:results}, we provide
EOM-CCSD results on the excitation energies (including a comparison with CIS
and a discussion of finite-size effects), the exciton binding energy, the
dispersion of excitons with nonzero momentum, and the exciton-phonon
interaction.  In section~\ref{sec:conclusions}, we summarize our results and
conclude with future directions.

\section{Theory}
\label{sec:theory}

Our periodic calculations are performed using a 
translational-symmetry-adapted single-particle basis,
\begin{equation}
    \phi_{\mu \vk}(\vr) = \sum_{\vT} \mathrm{e}^{\mathrm{i} \vk \cdot \vT} \tilde{\phi}_{\mu}(\vr-\vT) , 
\end{equation}
where $\tilde{\phi}_\mu(\vr)$ is
an atom-centered Gaussian orbital, $\vT$ is a lattice translation vector, and 
$\vk$ is a crystal momentum vector sampled from the first Brillouin zone.
A Hartree-Fock calculation in this basis produces crystalline orbitals (COs)
\begin{equation}
    \psi_{p\vk} (\vr) = \sum_{\mu} C_{\mu p}(\vk)\phi_{\mu\vk} (\vr), \label{eq:co}
\end{equation}
where $p$ is the band index and $C_{\mu p}(\vk)$ are the CO coefficients.  

The periodic CCSD energy and cluster amplitudes are determined by the usual
equations~\cite{coester1960,cizek1966,kummel1991,bartlett1981,bartlett2007,hirata2004},
\begin{align}
    E_0 & = \langle \Phi_0| \bar{H} | \Phi_0 \rangle, \label{eq:cc_energy_eq}                                                                                         \\
    0   & = \langle \Phi_{\p{i} }^{\p{a} }| \bar{H} | \Phi_0 \rangle, \label{eq:cc_amp_eq_t1}                                            \\
    0   & = \langle \Phi_{\p{i}  \p{j} }^{\p{a}  \p{b} } | \bar{H} | \Phi_0 \rangle, \label{eq:cc_amp_eq_t2}
\end{align}
where $\Phi_{\p{i} }^{\p{a} }$ and 
$\Phi_{\p{i}  \p{j} }^{\p{a}  \p{b} }$ are Slater
determinants with one and two electron-hole pairs,
indices $i, j, k, l$ denote occupied orbitals, and $a, b, c, d$ denote
virtual orbitals.
The similarity transformed Hamiltonian is given by
$\bar{H} \equiv e^{-T} H e^T$,
where $T = T_1+T_2$ is a momentum-conserving cluster operator with
\begin{align}
    T_1 & = \sum_{ai} \sideset{}{'}\sum_{\vk_a \vk_i} \cct{i}{a} 
        a_{\p{a}}^{\dagger} a_{\p{i}}, \label{eq:t1_def} \\
    T_2 & = \frac{1}{4} \sum_{abij} \sideset{}{'}\sum_{\vk_a \vk_b \vk_i \vk_j} 
        \cctt{i}{j}{a}{b} a_{\p{a}}^{\dagger} a_{\p{b}}^{\dagger}
        a_{\p{j}} a_{\p{i}}. \label{eq:t2_def}
\end{align}
The primed summations indicate momentum conservation.  

Excited states are accessed in coupled-cluster theory using the
equation-of-motion (EOM)
formalism~\cite{emrich1981,koch1990,stanton1993,kobayashi1994,bartlett2007,krylov2008},
which amounts to diagonalizing the effective Hamiltonian $\bar{H}$ in a
truncated space of excitations.  For neutral excitations considered in this
work, we use electronic excitation (EE) EOM-CCSD, where the diagonalization is
performed in the basis of 1-particle+1-hole and 2-particle+2-hole states.  We
study excitations with zero and nonzero momentum $\vq$ by using a
basis of determinants with the corresponding momentum.  The (right-hand) excited
state is therefore given by
\begin{equation}
|\Psi(\vq)\rangle = \left[R_1(\vq) + R_2(\vq)\right] e^{T} |\Phi_0\rangle
\end{equation}
with
\begin{align}
    R_1(\vq) & = \sum_{ai} \sideset{}{'}\sum_{\vk_a \vk_i} \ccr{i}{a} 
        a_{\p{a}}^{\dagger} a_{\p{i}}, \label{eq:r1_def} \\
    R_2(\vq) & = \frac{1}{4} \sum_{abij} \sideset{}{'}\sum_{\vk_a \vk_b \vk_i \vk_j} 
        \ccrr{i}{j}{a}{b} a_{\p{a}}^{\dagger} a_{\p{b}}^{\dagger}
        a_{\p{j}} a_{\p{i}}, \label{eq:r2_def}
\end{align}
where the primed summations indicate that the momenta sum to $\vq$.  The use of
translational symmetry in CCSD leads to a computational cost that scales like
$o^2 v^4 N_k^4 $, where $o$ is the number of occupied orbitals per unit cell,
$v$ is the number of virtual orbitals per unit cell, and $N_k$ is the number of
$k$-points sampled.  Further details about periodic Gaussian-based Hartree-Fock
and CCSD can be found in Ref.~\citenum{mcclain2017}.

\section{Computational Details}
\label{sec:comput}

We study eight semiconducting and insulating materials featuring
the diamond/zinc-blende crystal structure and the rock salt crystal structure.
The materials have a wide range of both direct and indirect band gaps and a
variety of ionic and covalent bonding.  The eight materials and the lattice
constants used are given in Tab.~\ref{tab:cis_finite_size}.
In all calculations, the Brillouin zone was sampled with 
a uniform Monkhorst-Pack mesh~\cite{monkhorst1976} of $N_k$ $k$-points that
includes the $\Gamma$ point.

Our calculations are performed with GTH pseudopotentials~\cite{goedecker1996,hartwigsen1998}, although
we perform some all-electron calculations for comparison.  For pseudopotential
calculations, we use the corresponding polarized double- and triple-zeta basis sets
DZVP and TZVP~\cite{vandevondele2005}.  For all-electron calculations, we
use a modification of the cc-pVDZ basis set presented in 
Ref.~\onlinecite{lorenz2012}, denoted AE-PVDZ.

The finite-size errors of periodic calculations are influenced by the
treatment of two-electron repulsion integrals (ERIs).  Many of these integrals
are formally divergent, due to the long-range nature of the Coulomb interaction;
however, these divergent ERIs enter into expressions for observables as integrable divergences,
producing well-defined results in the thermodynamic limit.
In order to avoid divergent ERIs, 
we calculate all atomic orbital ERIs as
\begin{equation}
(\mu\vk_\mu, \nu\vk_\nu | \kappa\vk_\kappa, \lambda\vk_\lambda)
= N_k^{-1} \int d\vr_1 \int d\vr_2 
    \frac{\rho_{\mu\nu}^{\vk_\mu\vk_\nu}(\vr_1) \rho_{\kappa\lambda}^{\vk_\kappa\vk_\lambda}(\vr_2) } {|\vr_1 - \vr_2 |} ,
\end{equation}
where the orbital-pair densities have had their net charge removed,
\begin{subequations}
\begin{align}
\rho_{\mu\nu}^{\vk_\mu\vk_\nu}(\vr) &= \phi_{\mu\vk_\mu}^*(\vr)\phi_{\nu\vk_\nu}(\vr) - \overline{\rho}_{\mu\nu} , \\
\overline{\rho}_{\mu\nu} &= \frac{1}{N_k \Omega} \int d\vr \phi_{\mu\vk_\mu}^*(\vr)\phi_{\nu\vk_\nu}(\vr),
\end{align}
\end{subequations}
and $\Omega$ is the volume of the unit cell.
The ERIs are then calculated using periodic Gaussian density fitting with an
even-tempered auxiliary basis as described in Ref.~\onlinecite{sun2017a}.
We note that the use of chargeless pair densities is equivalent to 
neglecting the $\vG=0$ component of the ERIs when
calculated in a plane-wave basis.
At the HF level, this treatment of ERIs produces an energy that converges
to the thermodynamic limit as $N_k^{-1/3}$, due to the exchange energy;
this can be corrected with a Madelung constant, leading to $N_k^{-1}$ convergence~\cite{oba2008,sundararaman2013}.
This particular finite-size behavior is also present in the occupied orbital energies $\varepsilon_i(\vk)$.

We use a closed-shell implementation of periodic EOM-CCSD
and study singlet excitations in this work, which are calculated by
Davidson diagonalization~\cite{davidson1975,hirao1982}.
The initial guess used in the iterative diagonalization is obtained from
dense diagonalization of the effective Hamiltonian in the single
excitation subspace.

All calculations were performed with the open-source PySCF software 
package~\cite{sun2018}.

\section{Results and Discussion}
\label{sec:results}

\subsection{Direct optical excitation energy}    
\label{sec:converg_k}

In this section, we study the performance of periodic EOM-CCSD on the
lowest-lying direct singlet excitation energies for the eight selected solids.
Such states are relevant for absorption spectroscopy where no momentum is transferred
during excitation.

In periodic electronic structure calculations, it is desirable to achieve convergence
with respect to the single-particle basis set, the level of correlation,
and Brillioun zone sampling. The first two categories have been widely
studied in molecular systems and the convergence behavior of the third one is
well understood at the mean-field level~\cite{carrier2007,spencer2008,sundararaman2013}. However, at the correlated
level, it is still an open question how to efficiently converge the Brillioun
zone sampling in order to reach the thermodynamic limit~\cite{gruneis2010,gruneis2011,booth2016,liao2016}.

\subsubsection{CIS}
\label{ssec:cis}

As a warm-up to EOM-CCSD, we first present results for periodic configuration
interaction with single excitations (CIS), which forms a minimal theory for
electronic excited states in the condensed phase.  
Importantly, the relatively low cost of CIS allows
us to study the convergence with respect to Brillouin zone sampling up to relatively
large $k$-point meshes (either $7^3$ or $8^3$).

The finite-size error of the CIS excitation energy can be analyzed approximately by considering it as
the difference between the energy of two determinants, the HF one $|\Phi_0\rangle$
and a single excitation $|\Phi_{i\vk}^{a\vk}\rangle$.  As discussed above, 
the energy of a single determinant calculated using our handling of ERIs exhibits
a finite-size error decaying like $N_k^{-1/3}$, but which can be removed by a Madelung
constant that depends only on the number of electrons in the unit cell.  Therefore,
the correction is identical for both states, and this leading-order error
cancels in the energy difference.
We thus posit that the CIS energy converges at least as fast as $N_k^{-1}$.

\begin{figure}[t!]
    \begin{center}
        \includegraphics[scale=0.85]{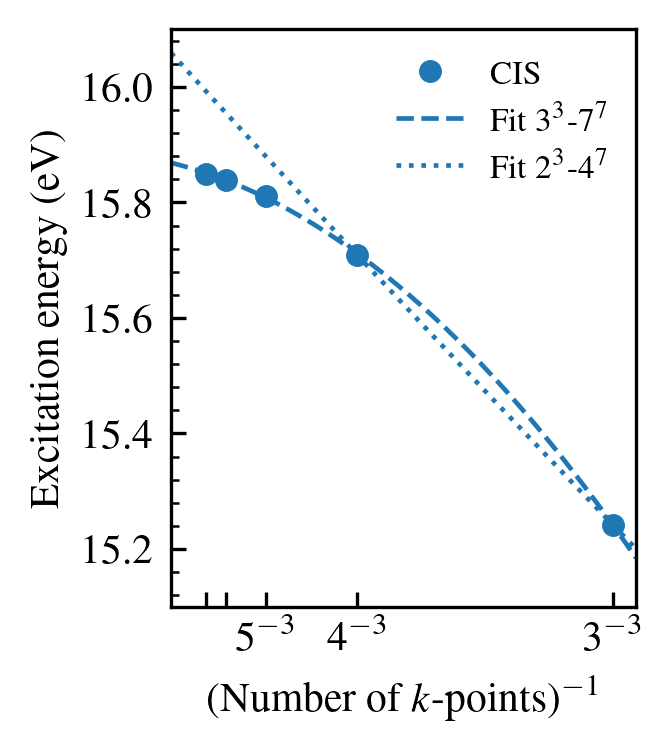}
        \caption{Excitation energy of LiF calculated with CIS using the DZVP basis set.  The dashed line
is a fit to results obtained with a number of $k$-points between $3^3$ and $7^3$.  The dotted line
includes only $2^3$, $3^3$, and $4^3$ $k$-points.
}
        \label{fig:lif_cis}
    \end{center}
\end{figure}

In Fig.~\ref{fig:lif_cis}, we show the excitation energy predicted by CIS for the LiF crystal
as a function of $N_k^{-1}$.  Clearly, the finite-size error decays at least
as fast as $N_k^{-1}$ and so we use the three-parameter fitting function
\begin{align}
    E(N_k) = E_{\infty} + a\, N_k^{-1} + b\, N_k^{-2}, \label{eq:extrap_n1n2}
\end{align}
in order to extrapolate to the thermodynamic limit ($N_k\rightarrow \infty$).
This fit is shown by the dashed line in Fig.~\ref{fig:lif_cis}, which includes results
between $3^3$ and $7^3$ $k$-points.

Using an all-electron double-zeta basis set, we performed CIS calculations on a
subset of our eight materials, in order to compare to previously published CIS
results from Lorenz et al~\cite{lorenz2012}.  By extrapolation of our
results obtained at large $k$-point meshes using the above form, we obtain converged excitation
energies (the ``AE-PVDZ/Conv'' column of Tab.~\ref{tab:cis_finite_size}) that are
within 0.1~eV of Ref.~\onlinecite{lorenz2012} for all materials studied.
However, we emphasize that the CIS excitation energies are larger than experiment
by 2~eV or more.

\begin{table}[!b]
    \caption{Singlet excitation energies (eV) from all-electron and pseudopotential-based periodic CIS}
    \label{tab:cis_finite_size}
    \begin{threeparttable}
        \begin{ruledtabular}
            \begin{tabular*}{0.48\textwidth}{l @{\extracolsep{\fill}} d{-1} d{-1} d{-1} d{-1} d{-1} d{-1}}
                \toprule
                System & \multicolumn{1}{c}{$a$ (\AA)} &\multicolumn{3}{c}{AE-PVDZ} & \multicolumn{2}{c}{DZVP}\\
                & & \multicolumn{1}{c}{234-Extrap\tnote{a}} & \multicolumn{1}{c}{Conv\tnote{b}} & \multicolumn{1}{c}{Ref. \citenum{lorenz2012}} & \multicolumn{1}{c}{234-Extrap\tnote{a}} & \multicolumn{1}{c}{Conv\tnote{b}} \\
                \cline{1-7}
                Diamond & 3.567 & 10.99    & 11.76    & 11.72 & 11.07   & 11.81  \\
                Si & 5.431 & 5.84     & 6.14     & 6.05  & 5.43    & 5.71  \\
                SiC     & 4.360 & 9.41     & 9.83     & 9.74  & 9.12    & 9.47  \\
                LiF     & 3.990 & 16.02    & 15.85    & 15.84 & 16.06   & 15.87  \\
                LiCl    & 5.130 &          &          &       & 11.04   & 10.89  \\
                MgO     & 4.213 & 12.00    & 11.91    & 11.94 & 11.69   & 11.66  \\
                BN      & 3.615 &          &          &       & 14.17   & 14.32  \\
                AlP     & 5.451 &          &          &       & 6.59    & 6.76 \\
                \bottomrule
            \end{tabular*}
        \end{ruledtabular}
        \begin{tablenotes}
            \item[a] Extrapolation based on results obtained with $2^3$, $3^3$, and $4^3$ $k$-point meshes 
            \item[b] Extrapolation based on results obtained with $3^3$ up to $7^3$ $k$-point meshes
        \end{tablenotes}
    \end{threeparttable}
\end{table}

Before moving on to our EOM-CCSD results, we use CIS to assess some of the
future approximations we will have to make.  In particular, we will only access
$k$-point meshes up to $4\times 4\times 4$ and we will use GTH pseudopotentials
and corresponding DZVP basis sets.  First, considering finite-size errors, we
re-fit the CIS data using the same form but excluding all $k$-point meshes
larger than $4\times 4\times 4$; for LiF, the result of this fit is shown as the
dotted line in Fig.~\ref{fig:lif_cis}.  Clearly, without larger $k$-point
meshes, this extrapolation predicts an excitation energy which is too high by
about 0.2~eV.  These limited extrapolation results are shown for all materials in the
``AE-PVDZ/234-Extrap'' column of Tab.~\ref{tab:cis_finite_size} and exhibit errors
of about $\pm 0.5$~eV.  Second, considering pseudopotential errors,
in the last two columns of Tab.~\ref{tab:cis_finite_size}, we show the
excitation energies calculated with GTH pseudopotentials.  In many cases, the pseudopotential
error is less than 0.1~eV; naturally, materials containing heavier second-row atoms such as Si
or Mg exhibit the largest errors, which are about 0.3~eV.

\subsubsection{EOM-CCSD}
\label{ssec:eomccsd}

We now move on to our results from periodic EOM-CCSD.  Again, due to the high
cost of these calculations, we utilized GTH pseudopotentials, sampled the
Brillouin zone with meshes up to $4\times 4\times 4$, and used basis set
corrections.  In particular, our primary calculations were based on a HF
reference obtained with the full DZVP basis set; subsequent CCSD and EOM-CCSD
calculations then employed frozen virtual orbitals, typically correlating 4 lowest
virtual bands.  The results of these calculations were used to extrapolate to
the thermodynamic limit using the same empirical formula as described above (Eq.~\ref{eq:extrap_n1n2}).
These results are given in the ``234-Extrap'' column of
Tab.~\ref{tab:ee_finite_size}.  To these values, we then added two basis set
corrections, $\Delta_\mathrm{frz}$ and $\Delta$TZ; $\Delta_\mathrm{frz}$ is the
energy difference between complete and frozen-orbital DZVP calculations with a
$3\times 3\times 3$ $k$-point mesh; $\Delta$TZ is the energy difference between
TZVP and DZVP calculations with a $2\times 2\times 2$ $k$-point mesh.
Whereas $\Delta_\mathrm{frz}$ is typically between 0.2
and 0.8~eV, $\Delta$TZ is less than 0.1~eV.

\begin{table}[!t]
    \caption{Singlet excitation energies (eV) from periodic EOM-CCSD}\label{tab:ee_finite_size}
    \begin{threeparttable}
        \begin{ruledtabular}
            \begin{tabular*}{0.48\textwidth}{l @{\extracolsep{\fill}} d{-1} d{-1} d{-1} d{-1} c}
                \toprule
                System & \multicolumn{1}{c}{234-Extrap\tnote{a}} & \multicolumn{1}{c}{$\Delta_{\mathrm{frz}}$\tnote{b}} & \multicolumn{1}{c}{$\Delta$TZ\tnote{c}} & \multicolumn{1}{c}{Final} & \multicolumn{1}{c}{Expt.} \\
                \hline
                Diamond & 7.70  & -0.18 & -0.05 & 7.47  & 7.3~\cite{phillip1964} \\
                Si      & 3.96  & -0.37 & -0.07 & 3.52  & 3.4~\cite{lautenschlager1987}  \\
                SiC     & 6.53  & -0.19 & -0.08 & 6.27  & 6.0~\cite{logothetidis1996}  \\
                LiF     & 14.29 & -0.82 & +0.01\tnote{d} & 13.48 & 12.7~\cite{rao1975}, 13.68~\cite{abbamonte2008} \\
                LiCl    & 9.62  & -0.27 & -0.07 & 9.29  & 8.9~\cite{eby1959} \\
                MgO     & 8.55  & -0.19 & -0.07 & 8.29  & 7.6~\cite{roessler1967} \\
                BN      & 11.48 & -0.35 & -0.02 & 11.11 & 11~\cite{tararan2018} \\ 
                AlP     & 4.97  & -0.42 & -0.07 & 4.48  & 4.6~\cite{hwang2014,wing2019} \\
                \hline
                MSE & & & & 0.24 & \\
                MAE & & & & 0.27 & \\
                \bottomrule
            \end{tabular*}
        \end{ruledtabular}
        \begin{tablenotes}
            \item[a] Extrapolation based on frozen-virtual DZVP results obtained with $2^3$, $3^3$, and $4^3$ $k$-point meshes
            \item[b] $\Delta_\mathrm{frz}$ is the energy difference between complete DZVP and frozen-virtual DZVP on a 3$\times$3$\times$3 $k$-point grid 
            \item[c] $\Delta$TZ is the energy difference between TZVP and DZVP on a 2$\times$2$\times$2 $k$-point grid
            \item[d] For LiF, the TZVP basis set has severe linear dependencies and was modified by doubling the exponents of the two most diffuse s- and p-type primitive Gaussian functions of Li (all basis functions of F are unchanged)
        \end{tablenotes}
    \end{threeparttable}
\end{table}

\begin{figure}[t!]
    \begin{center}
        \includegraphics[scale=0.85]{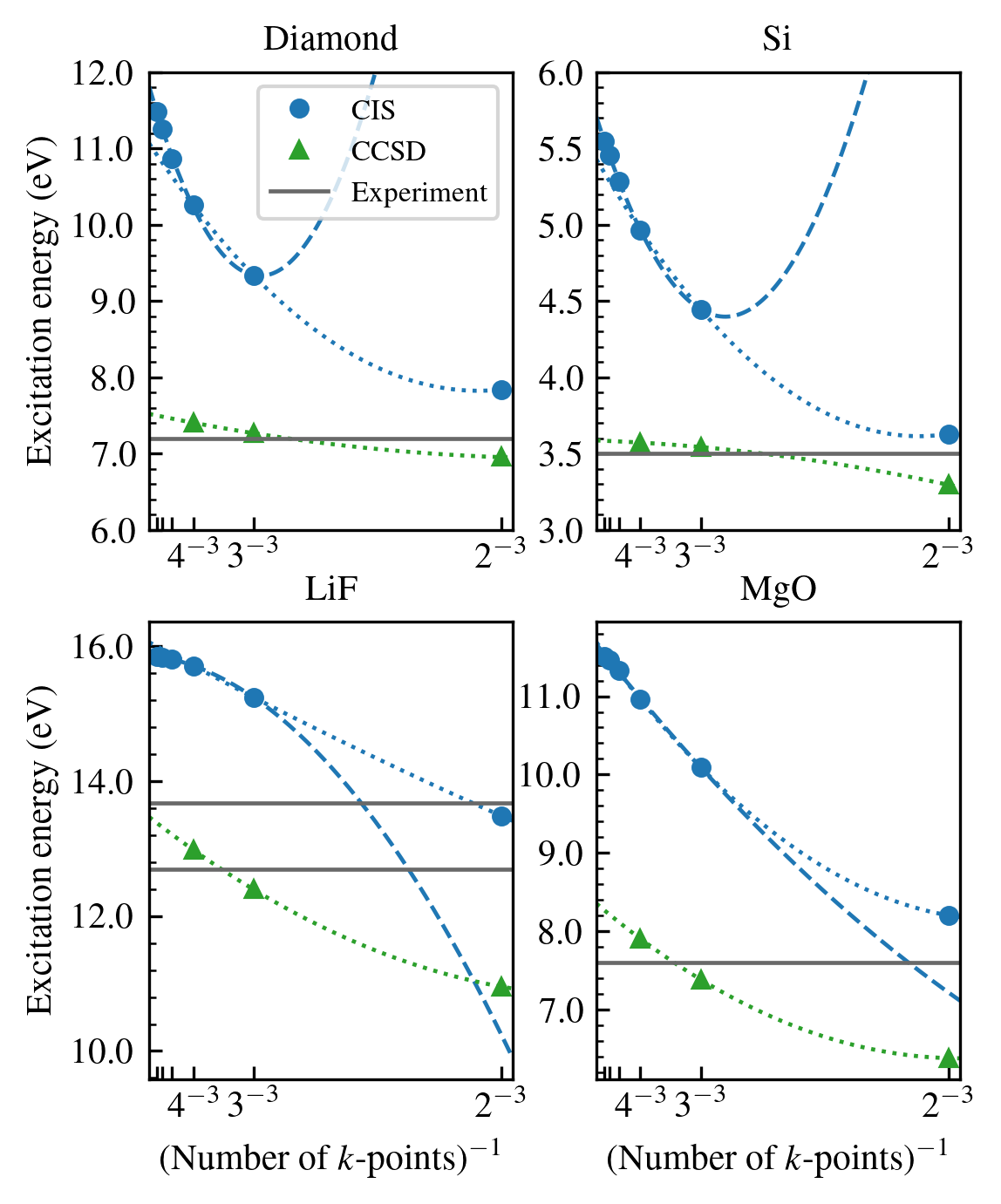}
        \caption{Excitation energy of diamond, Si, LiF, and MgO calculated with
CIS and EOM-CCSD.  Results are extrapolated to $N_k \rightarrow \infty$ assuming
an error with finite-size scaling as described in the text. The dashed lines
are fitted based on results obtained with a number of $k$-points between $3^3$ and $7^3$.  
The dotted lines include only $2^3$, $3^3$, and $4^3$ $k$-points. A variety of
experimental results are indicated by solid lines. }
        \label{fig:ee_vs_kpts}
    \end{center}
\end{figure}

Basis-set corrected excitation energies as a function of $N_k^{-1}$ are shown in
\cref{fig:ee_vs_kpts} for diamond, Si, LiF, and MgO.  Our final values for all
eight materials are given in the ``Final'' column of
Tab.~\ref{tab:ee_finite_size} and compared to experiment.  
Unsurprisingly, EOM-CCSD is a massive
improvement over CIS; for the eight solids studied here, EOM-CCSD predicts 
excitation energies with a mean signed error (MSE) of 0.24~eV and a mean
absolute error (MAE) of 0.27~eV.  

A few results in Tab.~\ref{tab:ee_finite_size} are noteworthy. First,
we note that the excitation energy of cubic BN is frequently reported
as 6.4~eV, which is almost 5~eV lower than the EOM-CCSD prediction.
However, a GW/BSE calculation reported in 2004~\cite{satta2004} also found
a value of around 11~eV and proposed a reinterpretation of the experimental
data.  Indeed, a joint theory-experiment paper published in 2018~\cite{tararan2018} attributed the
lower energy absorption features to a combination of defects and domains
of hexagonal BN, further supporting a direct excitation energy of about
11~eV.  Second, the excitation energy of AlP has frequently been reported
as 3.6~eV, about 1~eV below the EOM-CCSD prediction.  A 2019
publication reporting the results of TDDFT and GW/BSE calculations~\cite{wing2019} suggested
that the 3.6~eV feature seen in experimental spectra is due to the
\textit{indirect} transition of AlP.  Experimental spectra show a much stronger
peak at around 4.6~eV~\cite{hwang2014}, which is the likely value of the first direct excitation
energy.

The two materials with the largest error are LiF and MgO.  Whereas absorption
spectra of LiF typically show a narrow peak at about 12.7~eV~\cite{rao1975} (leading to an
overprediction of 0.8~eV), inelastic X-ray scattering data is consistent with a
value of 13.68~eV~\cite{abbamonte2008}.  For MgO, EOM-CCSD overpredicts the excitation energy by
about 0.7~eV.  This tendency to overestimate excitation energies is the same as
the one typically observed in molecules and could potentially be addressed via
inclusion of triple excitations.  However, we also emphasize that our
calculations include no information about finite-temperature or exciton-phonon
effects, which are expected to contribute to a reduction in the purely
electronic excitation energy~\cite{marini2008,giustino2010,antonius2014,ponce2014,zacharias2015}.

\subsection{Exciton binding energy}
\label{sec:exciton_binding}

We now consider the exciton binding energy, defined as the difference between
the fundamental band gap and the first neutral excitation energy (i.e.~the
optical gap).  Within periodic coupled-cluster theory, the band gap is obtained
from the calculation of the ionization potential (IP-EOM-CCSD) and the electron
affinity (EA-EOM-CCSD), as described in
Ref.~\onlinecite{mcclain2017}.  Here, our IP/EA-EOM-CCSD
calculations are basis-set-corrected in the same way as described for our
EOM-CCSD calculations.  

\begin{figure}[t!]
    \begin{center}
        \includegraphics[scale=0.85]{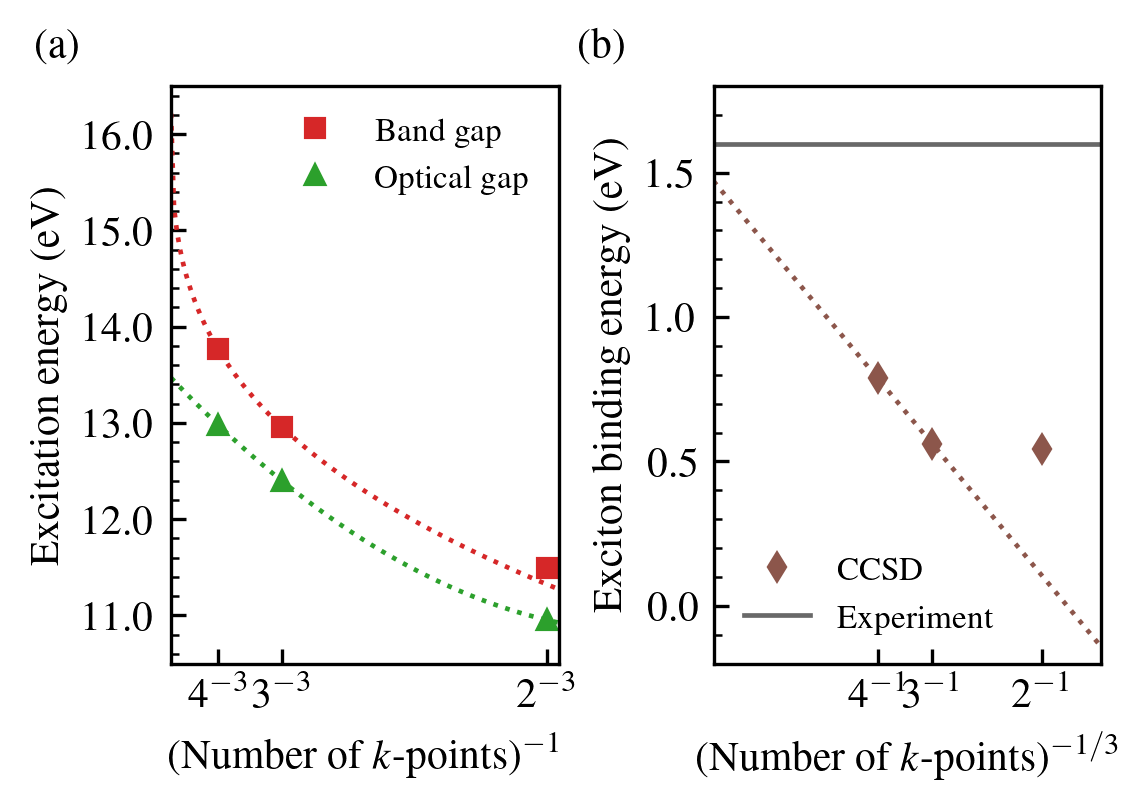}
        \caption{Fundamental band gap and optical gap (a) and exciton binding
energy (b) of LiF from periodic EOM-CCSD.  Due to the behavior of
IP/EA-EOM-CCSD, the fundamental band gap (red squares) and exciton binding energy (brown diamonds) are
extrapolated to $N_k \rightarrow \infty$ assuming a finite-size scaling of the
form $E_{N_k} = E_{\infty} + a\,N_k^{-1/3}$. }
        \label{fig:exciton_binding}
    \end{center}
\end{figure}

We focus on LiF, which is a direct-gap ionic insulator with a concomitantly
large exciton binding energy.  The minimum band gap occurs at the $\Gamma$
point, which is where we calculate the IP and EA.  In
Fig.~\ref{fig:exciton_binding}(a), we show the fundamental (IP/EA) gap and the
optical (EE) gap, as a function of the number of $k$-points sampled in the
Brillouin zone.  Clearly the fundamental gap is larger than the optical gap such
that the exciton binding energy is positive, as expected.  The exciton binding
energy is plotted in Fig.~\ref{fig:exciton_binding}(b), and seen to be around
0.8~eV with a $4\times 4\times 4$ $k$-point mesh.  Unlike neutral excitation
energies, which conserve particle number, the IP and EA are charged excitation
energies corresponding to a change in particle number.  
In particular, the same approximation considered above for CIS, i.e.~the use of a single determinant,
leads to Koopmans' approximation to the ionization potential,
IP$\approx -\varepsilon_i$; as discussed above, occupied orbital
energies converge slowly with $N_k$ when the ERIs are handled as described
in section~\ref{sec:comput}.  Therefore, as shown in
Fig.~\ref{fig:exciton_binding}(b), we fit the exciton binding energies
calculated with $3\times 3\times 3$ and $4\times 4\times 4$ grids to the form
$E(N_k) = E_\infty + a N_k^{-1/3}$, in order to extrapolate to the thermodynamic
limit.  Doing so gives 1.47~eV, which is in good agreement with the
experimental value of 1.6~eV~\cite{roessler1967}.  We note that if we separately
extrapolate the band gap and the optical gap and take the difference, we get a
larger value of 2.74~eV.

\subsection{Exciton dispersion}
\label{sec:ee_vs_kshift}

\begin{figure}[b!]
    \begin{center}
        \includegraphics[scale=0.85]{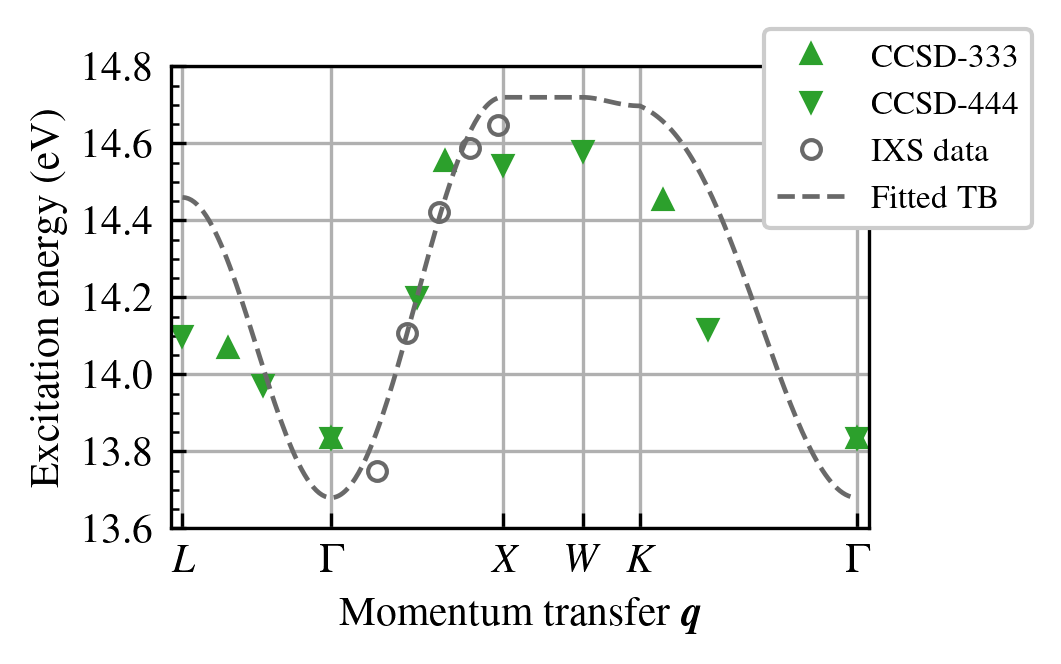}
        \caption{Exciton dispersion of LiF. Periodic EOM-CCSD data are obtained
with different $k$-point meshes in order to access more values of the momentum
transfer $\vq$.  Rigid shifts were applied, as described in the text.  Also
shown are the experimental inelastic X-ray scattering (IXS) data from
Ref.~\citenum{abbamonte2008} and the tight-binding (TB) model fitted to that
data.}
        \label{fig:ee_vs_kshift}
    \end{center}
\end{figure}

Although optical absorption spectroscopy and modern theoretical approaches 
have primarily focused on excitons with $\vq = 0 $, it is important to also
consider excitons that carry a finite momentum $\vq$. 
For example, electron-hole pairs with finite momentum are realized
in many indirect semiconductors and are also important for a quantitative
modeling of the exciton-phonon interaction, excitonic dynamics, and
radiative lifetimes.

The simulation of excitons with finite-momentum is straightforward in
EOM-CCSD, and simply requires that the involved crystal momenta sum to $\vq$
in \cref{eq:r1_def,eq:r2_def}.  The EOM Hamiltonian is block-diagonal with
respect to the exciton momentum and thus all accessible momenta can be studied
independentaly and calculated in parallel.
Because the exciton momentum $\vq$ corresponds to a momentum difference,
a periodic calculation only has access to values $\vq = \vk - \vk^\prime$,
where $\vk$ are momenta from the $k$-point mesh.
Therefore, different $k$-point meshes can be utilized in order to access
different values of the exciton momentum $\vq$, albeit with an impact
on the finite-size error.

Again we focus on LiF, for which we show the EOM-CCSD exciton dispersion
in Fig.~\ref{fig:ee_vs_kshift}.  
Our results are compared to inelestic X-ray scattering (IXS) spectroscopy
measurements performed by Abbamonte et al.~\cite{abbamonte2008} (open circles).
We utilize $3\times 3\times 3$ $k$-point mesh (up-pointing triangles) and $4\times 4\times 4$
$k$-point mesh (down-pointing triangles) in order to access more momenta $\vq$ in the Brillouin zone.
The $3\times 3\times 3$ values were rigidly shifted in order to
achieve agreement at the $\Gamma$ point and subsequently, both dispersion data were 
rigidly shifted in order to place the exciton band center at 14.2~eV,
as was observed experimentally.

From their experimental data along the $\Gamma-X$ line, those authors
paramaterized a tight-binding model (with band center 14.2~eV and
nearest-neighbor transfer integral $-0.065$~eV), which we have extended to the
entire Brillouin zone for comparison (dashed line).  We can see that the
EOM-CCSD results are in good agreement with the IXS data, with an error less
than 0.2 eV along the $\Gamma \rightarrow X$ direction.  Our largest discrepancy
is at the $L$ point, although we emphasize that experimental data is unavailable
at that momentum and the disagreement may indicate a failure of the simple
tight-binding model.

\subsection{Exciton-phonon interaction}
\label{sec:ee_vs_volume}

\begin{figure}[t!]
    \begin{center}
        \includegraphics[scale=0.85]{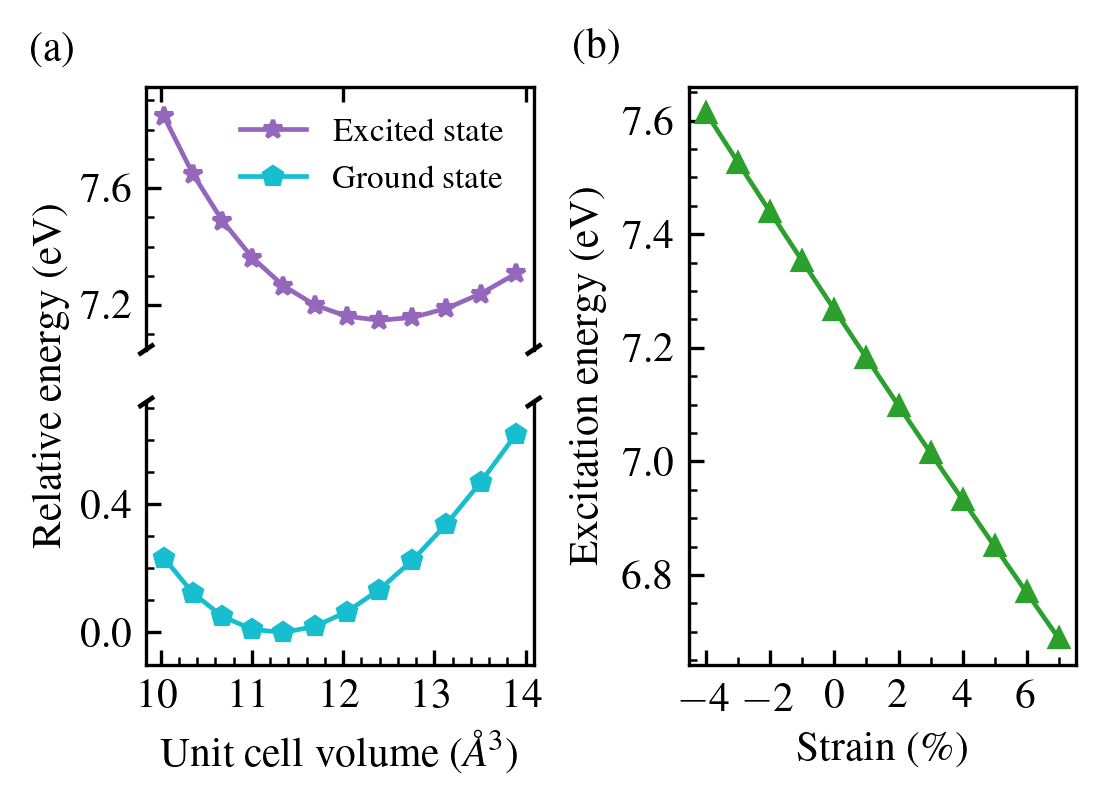}
        \caption{Ground- and excited-state potential energy surface of diamond
as a function of the unit cell volume (a) and the behavior of the excitation
energy as a function of hydrostatic strain (b).  Results are obtained from CCSD
and EOM-CCSD with the DZVP basis set and a $3\times 3\times 3$ sampling of the
Brillouin zone.}
        \label{fig:diamond_pes_ee_vs_vol}
    \end{center}
\end{figure}

Finally, we consider the behavior of excitation energies as a function of
lattice strain, as predicted by EOM-CCSD.  In
Fig.~\ref{fig:diamond_pes_ee_vs_vol}(a), we show the ground-state and first
excited-state potential energy surfaces of diamond, associated with hydrostatic
strain, i.e.~isotropic variation of the unit cell.  While the ground-state
energy minimum occurs at a lattice constant of 3.567~\AA~(which is fortuitously
the exact experimental value~\cite{haas2009,kittel1996}), the excited state has a minimum which
is shifted to a larger lattice constant of 3.674~\AA.  

The behavior of the excitation energy as a function of lattice strain can be
used to determine properties of the exciton-phonon interaction.  In particular,
the interaction with acoustic phonons can be modeled within the deformation
potential approximation for the change in the excitation energy~\cite{bardeen1950,herring1956}, $\Delta
E_\mathrm{X} = 3 D \varepsilon$, where $D$ is the hydrostatic deformation
potential, $3\varepsilon$ is the trace of the strain tensor, $\varepsilon
= (a-a_0)/a_0$ is the relative strain, $a$ is the strained lattice constant, and $a_0$ is the unstrained
lattice constant.  Defined in this way, our calculations predict the excitonic hydrostatic
deformation potential of diamond to be $D=-2.84$~eV.  Repeating the same
procedure for MgO, we predict a larger $D=-11.73$~eV, indicating a stronger
exciton-phonon interaction than in diamond.  Experimentally determined
deformation potentials for exciton-phonon interactions are sporadic in the
literature and we consider a direct comparison on a wider variety of materials
to be a topic for future work.

\section{Conclusions}
\label{sec:conclusions}

To summarize, we have presented the results of periodic EOM-CCSD for various neutral excited-state properties of
semiconductors and insulators.  The method has shown promising results for optical excitation energies,
exciton binding energies, exciton dispersion relations, and exciton-phonon interaction energies.
Collectively, these results demonstrate that EOM-CCSD is a promising and tractable approach
for the study of excited-state properties of solids.

While we have attempted to address finite-size errors, arising from incomplete
sampling of the Brillouin zone, the high cost of EOM-CCSD precludes a definitive
extrapolation.  Future work will be focused on both analytical and numerical
exploration of finite-size errors and strategies for amelioration, such as those
that have been developed for ground-state CCSD~\cite{gruneis2011,booth2016,liao2016}.  In a similar
direction, we are exploring the use of approximations to EOM-CCSD with reduced
scaling~\cite{goings2014}, which will enable the study of simple solids with more $k$-points or
the study of solids with more complex unit cells.
Additional future work is targeted at the study of simple metals (where finite-order
perturbation theory breaks down~\cite{shepherd2013}), the study of exciton-phonon interactions beyond
the deformation potential approximation, and the efficient calculation of optical spectra.

\begin{acknowledgments}
X.W.~thanks Dr.~Varun Rishi for helpful discussions. 
This work was supported in part by the National Science Foundation under 
Grant No.~CHE-1848369.
All calculations were performed using resources provided by the Flatiron
Institute.  
The Flatiron Institute is a
division of the Simons Foundation.
\end{acknowledgments}

\raggedbottom

\end{document}